%% file: kasper_paper_final.tex
\documentclass[sigconf]{acmart}
\usepackage{tikz}
\AtBeginDocument{%
  \providecommand\BibTeX{{%
    \normalfont B\kern-0.5em{\scshape i\kern-0.25em b}\kern-0.8em\TeX}}}

\setcopyright{acmcopyright}
\copyrightyear{2022}
\acmYear{2022}
\acmConference[The 2022 SIGIR Workshop On eCommerce]{Make sure to enter the correct}{July 15 2022}{Madrid - Spain}
\setcopyright{rightsretained}
\acmConference[SIGIR eCom'22]{ACM SIGIR Workshop on eCommerce}{July 15, 2022}{ Madrid, Spain}
\acmYear{2022}
\copyrightyear{2022}
\makeatletter
\renewcommand\@formatdoi[1]{\ignorespaces}
\makeatother
\acmISBN{}
\begin{document}

\title{Contrastive Learning for Interactive Recommendation in Fashion}


\author{Karin Sevegnani, Arjun Seshadri, Tian Wang, Anurag Beniwal, Julian McAuley, Alan Lu, Gerard Medioni}
\affiliation{%
 \institution{Amazon}
 \city{San Francisco}
 \country{USA}}

\email{karin.sevegnani@hw.ac.uk, {sesarjun, wangtan, beanurag, jumcaule, alalu, medioni}@amazon.com}
    
% \author{Karin Sevegnani}
% \email{karin.sevegnani@hw.ac.uk}

% \author{Arjun Seshadri}
% \email{sesarjun@amazon.com}

% \author{Tian Wang}
% \email{wangtan@amazon.com}
  
% \author{Anurag Beniwal}
% \email{beanurag@amazon.com}

% \author{Julian McAuley}
% \email{jumcaule@amazon.com}
 
% \author{Alan Lu}
% \email{alalu@amazon.com}
 
% \author{Gerard Medioni}
% \email{medioni@amazon.com}

% \affiliation{%
%  \institution{Amazon}
%  \city{San Francisco}

\renewcommand{\shortauthors}{Sevegnani et al.}

\begin{abstract}
Recommender systems and search are both indispensable in facilitating personalization and ease of browsing in online fashion platforms. However, the two tools often operate independently, failing to combine the strengths of recommender systems to accurately capture user tastes with search systems' ability to process user queries. We propose a novel remedy to this problem by automatically recommending personalized fashion items based on a user-provided text request.
Our proposed model, WhisperLite, uses contrastive learning to capture user intent from natural language text and improves the recommendation quality of fashion products.
WhisperLite combines the strength of CLIP embeddings with additional neural network layers for personalization, and is trained using a composite loss function based on binary cross entropy and contrastive loss.
The model demonstrates a significant improvement in offline recommendation retrieval metrics when tested on a real-world dataset collected from an online retail fashion store, as well as widely used open-source datasets in different e-commerce domains, such as restaurants, movies and TV shows, clothing and shoe reviews.
We additionally conduct a user study that captures user judgements on the relevance of the model's recommended items, confirming the relevancy of WhisperLite's recommendations in an online setting.
\end{abstract}

\maketitle

\section{Introduction}

Recommendation systems are at the core of e-commerce websites such as Amazon, Netflix, and Ebay.
These systems recommend products based on the user's historical preferences and purchases \cite{zhang2019deep, ricci2015recommender}.
The goal is to recommend products to users that they might like and improve the experience to make it enjoyable and satisfactory.

\begin{figure}[H]
\small
\begin{tabular}{|p{5cm}p{1.6cm}|}
\hline
\vspace{0.5cm}\textbf{Text request:} No dresses, skirts, shorts or heels please. I like Shoes like Dr. Martens, casual low to no heel. & \vspace{0.1cm}\includegraphics[width=1.5cm]{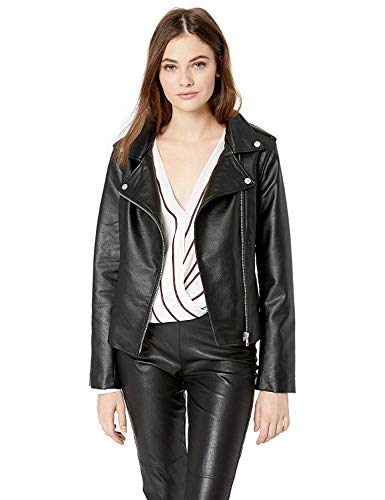}\vspace{0.1cm} \\
\hline
\vspace{0.5cm}\textbf{Text request:} Show me some spring and summer wear. I'm a mom so please no crop tops  & \vspace{0.1cm}\includegraphics[width=1.5cm]{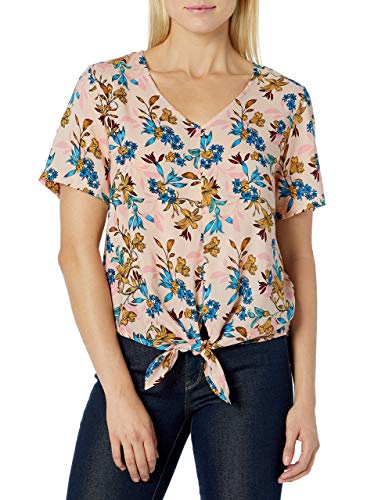}\vspace{0.1cm} \\
\hline
\end{tabular}
\centering
\caption{Examples of personalized text requests written by the user browsing a clothing retail online shop on the left, and an accordingly recommended item on the right.}
\label{fig:example}
\end{figure}

The goal of this work is to recommend fashion items focusing specifically on the user's free-form text request provided to a fashion stylist while using a personalized styling subscription service.
%In order to receive product recommendations that are aligned with their ideas, the customer types their request in a text note.%
In our problem setting, we only rely on the text request provided by the user, and do not assume knowledge of user  click-stream information or prior purchases. Although prior purchases are often used to model user preferences \cite{schafer2007collaborative}, the presence of such information cannot be assumed for so-called "cold start" users that are using a service for the first time and have no historical purchase or click stream information associated with them. Moreover, text requests often reveal more nuanced aspects of user preferences at the time of placing a styling request, making our work essential to improving existing recommender systems.

Our primary dataset consists of three parts: text requests from users of a personalized styling service, items chosen by a human stylist for the user based on those text requests, and items the user chooses to try from the stylist's recommended items.
% This work focuses on generating recommendations over a product catalogue in the fashion domain starting by a request note expressed by the customer.
Figure \ref{fig:example} shows two stylized examples, where users communicate a preference towards specific clothing items through a text request, and professional fashion stylists respond with a product recommendation pertaining to the customer request.

Our modeling task is challenging on several levels. First, text requests are more elaborate than a search query, and contain both useful contextual information that captures high-level user preferences and less helpful information that is irrelevant for retrieving fashion items e.g. ~"\textit{I am a school teacher and I don't wear yoga pants}."
Second, each item in the recommended set of items might only meet a subset of criteria specified by the user in the text request, creating opportunities for model misattribution and overfitting. Even the recommended set of items as a whole might not cover all of the criteria mentioned by the user. Third, there are items in the dataset selected by stylists that the user could potentially like despite the items not having the exact attributes requested by the user, further exacerbating the misattribution problem. Finally, there could be items that the user selected despite being irrelevant to the text request, or sometimes even contradictory to their initial text request. Each of these factors introduce noise in the dataset labels, and together create a challenging modeling problem.

%1) the language used by customers is quite noisy and can contain typos, similar to the comments found on social media websites, 2) the text request from the users do not always contain specific needs and style requests, as it is common for this type of information to be omitted, 3) the recommendation of fashion items is a very subjective task that requires a lot of information on the customer and on the event/surroundings, and finally 4) we approach the task leveraging textual features only rather than using a multi-modal feature approach, due to the type of information contained in the considered datasets.

% Additionally, the solution we propose will further be employed to deal with customer needs in a real fashion store.
% Ideally, users will send their request in advance and will receive their recommended pieces in the fitting room as soon as they enter the shop.

The contributions of this paper can be summarized as follows:
\begin{itemize}
    \item Defining and addressing a real-world recommendation task for online fashion retail %that uses elaborate text requests from users to recommend new items% 
    in order to improve the user experience (Section \ref{sec:formulation}),
    \item Validating the multi-domain generalization of the system offline using a real-world dataset and three other publicly available data sources (Section \ref{sec:results}),
    \item Validating the online performance of the system through a user study on Amazon Mechanical Turk (Section \ref{sec:human_eval}).
\end{itemize}

\section{Related Work} \label{sec:lit_rev}
The literature of recommender systems can be broadly segmented into collaborative filtering \cite{su2009survey, koren2015advances} and content-based methods \cite{pazzani2007content, balabanovic1997fab}.
While collaborative filtering techniques only exploit past user purchases and product interactions to generate new recommendations \cite{herlocker2000explaining}, content-based methods rely on information about the users or the items \cite{basu1998recommendation}. Recommendation systems for fashion retail e-commerce, the domain of our work, frequently fall into the content-based category, and are often paired with images and textual descriptions of the products \cite{chakraborty2021fashion, kang2017visually, chen2019personalized, yin2019enhancing, hwangbo2018recommendation, tu2010intelligent, ajmani2013ontology}. Our work is similar in being content-based to other works in fashion retail e-commerce, as it leverages textual user requests as the primary representation of the user. Within the broader literature of recommendation systems, our work most closely aligns with session-based recommendation approaches, which focus on predicting the user's immediate next actions using explicit feedback of the customer during the current session. % only the behavior of the customer during the current session. 
In our setting, the user's text request serves as the representation of the customer's preferences in the current session. We refer the reader to \cite{ludewig2018evaluation}, which provides a comprehensive review of the session-based recommendation literature.

\section{Methodology}
\subsection{Problem Formulation} \label{sec:formulation}
Our work approaches the recommendation problem by trying to build affinity between a natural language text query expressed by the user and an item description.
The natural language query $\textbf{x}=(x_1, x_2, ..., x_N)$ can consist of several sentences, and the item description $\textbf{z}=(z_1, z_2, ..., z_M)$ can include several item attributes and characteristics, where $N$ and $M$ are the number of tokens in the request and the item description, respectively.
The similarity between the request $\textbf{x}$ and the description $\textbf{z}$ is determined by the cosine distance between the two. Once similarity scores are computed, the various similarities are then ranked and used to perform a retrieval task. The success of the task is evaluated with different retrieval metrics, which we describe in Section \ref{sec:data_metrics}.

% \begin{align}
% \hat{\textbf{y}} = cos(\theta)   \\  
% \theta = (\hat{\textbf{x}}, \hat{\textbf{z}}) 
% \end{align}

% where $\hat{\textbf{x}}$ and $\hat{\textbf{z}}$ represent feature vectors for the text request and the product description, respectively.

We present two solutions to our problem setting: 
\begin{itemize}
    \item The Whisper approach: training a model that refines word embeddings for text notes and item descriptions in order to closely pair requests and products in the word embeddings space and treat them as a match for recommendation (Section \ref{sec:kasper}); and
    \item The WhisperLite approach: derive embeddings from a pre-trained system and train a model only to learn which items to recommend using contrastive loss (Section \ref{sec:klite}).
\end{itemize}

\subsection{Whisper} \label{sec:kasper}

\begin{figure}
    \resizebox{\linewidth}{!}{\input{arch.tex}}
    \caption{Architecture of the Whisper model. The dashed lines mean that the weights for XLNet are frozen when computing the embeddings for the product description.}
    \label{fig:Whisper}
\end{figure}
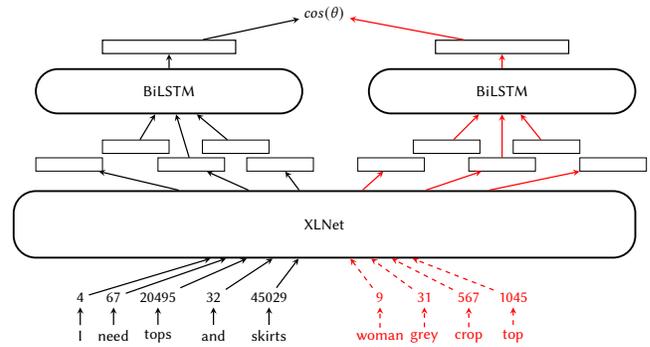
    
%     \begin{subfigure*}[h]
%     \centering
%     \input wlite.tex
%     \caption{Architecture of the Whisper model. The dashed lines mean that the weights for XLNet are frozen when computing the embeddings for the product description.}
%     \label{fig:WLite}
% \end{subfigure*}
% \end{figure*}

% \begin{figure}
% \centering
% \begin{minipage}{.1\textwidth}
%   \small
%   \input arch.tex
%   \captionof{figure}{A figure}
%   \label{fig:test1}
% \end{minipage}%
% \begin{minipage}{.1\textwidth}
%   \input wlite.tex
%   \captionof{figure}{Another figure}
%   \label{fig:test2}
% \end{minipage}
% \end{figure}

Large pre-trained transformers such as BERT \cite{devlin2018bert}, RoBERTa \cite{liu2019roberta}, or XLNet \cite{yang2019xlnet} have  proven particularly effective at the generation of contextual word embeddings.
The architecture of the Whisper model is planned and developed in a modular way.
We initially leverage one of the state-of-the-art large language models for generating word embeddings.
XLNet is the transformer of our choice, since it has been proven to outperform BERT on a variety of tasks \cite{malon2021overcoming, yang2019xlnet}.

In order to prepare the initial text requests and item descriptions, they are first split into tokens and assigned input ids based on the model tokenizer's built-in vocabulary.
Moreover, we perform a keyword search for each text request to highlight categorical tokens and strengthen the division of the text requests into the main clothing categories.
%If present, such categorical keywords are appended to the user's request through a \texttt{<sep>} token.

The pre-processed inputs are then passed into XLNet, which generates contextual word embeddings in an auto-regressive way. 
The embeddings of the words are one-dimensional tensors retrieved through the last hidden state of the language model.
Then, the embeddings are fed into a bidirectional LSTM, which produces one feature vector for each text request received as input.
A second bi-LSTM is used in order to generate different feature vectors for the product descriptions in a two-tower fashion \cite{yang2020mixed}, since the semantics of user requests are very different from those of product descriptions.

As shown in Figure \ref{fig:Whisper} of the Whisper architecture, the transformer's task is to generate contextual word embeddings for both the text request and the item descriptions.
The two are then separated and passed as input in the two LSTMs as mentioned above, producing the two corresponding feature vectors.
The model is trained using Cosine Embedding Loss 
\begin{align}
loss(x,y) = \begin{cases} 1-cos(x_{1}, x_{2}),   \text{if $y=1$} \\
max(0, cos(x_{1}, x_{2})),  \text{if $y=-1$ }
\end{cases}    
\end{align}

\subsection{WhisperLite} \label{sec:klite}

% \begin{figure}[h]
%     \resizebox{\linewidth}{!}{\input{wlite.tex}}
%     \caption{Architecture of the WhisperLite model. The weights of CLIP are frozen when computing the embeddings for both the text note and the product description.}
%     \label{fig:WLite}
% \end{figure}

WhisperLite is a 
%much 
lighter version of the Whisper model described in the previous section, where the text embeddings are calculated through the pre-trained CLIP text-only model \cite{radford2021learning} as opposed to XLNet.
Moreover, WhisperLite uses MLPs instead of bi-LSTM layers, and is trained using a customized loss function, combining binary cross entropy loss for classification 
\begin{align} \label{eq_bce}
\mathcal{L}_{BCE}=-(ylog(\hat{y})+(1-y)log(1-\hat{y}))
\end{align}
with a contrastive loss.
In \eqref{eq_bce}, $\hat{y}$ is the predicted probability of observation $o$ being in class $c$, and $y$ represents whether $o$ actually belongs in $c$.

Additionally, we exchanged the initial cosine similarity to the dot product between the request and the product description vectors since the two metrics are comparable \cite{luo2018cosine, naveen2021abstractive}.
The CLIP parameters are held frozen, and only the MLPs are trained, significantly reducing the training time of WhisperLite.

\section{Experiments}
\subsection{Compared models}
We consider two baseline models to compare with our proposed approach.
\begin{itemize}
    \item A \textbf{random} baseline, where the \texttt{(request,item)} pairs are assigned a random ranking.
    \item \textbf{OkapiBM25}, a function that ranks the item description based on its estimated IDF to the text request that the user provides.
\end{itemize}

Since the task addressed in this project deals with query-to-item recommendations without additional information about the user, to the best of our knowledge there are no other baselines that fit the problem exactly. As a result, our current baselines for comparison are elementary, and adapting existing work to suit our problem setting is an important direction for future work.

\subsection{Datasets and Evaluation} \label{sec:data_metrics}
% \begin{table*}[h!]
% \centering
% \small
% \begin{tabular}{|l|l|l|l|l|}
% \hline
%                   & Number  & Unique & \begin{tabular}[c]{@{}l@{}}Sentiment\\ (pos-neutr-neg)\end{tabular}                         & \begin{tabular}[c]{@{}l@{}}Word length\\ (min-max-avg-50th perc.-90th perc.)\end{tabular} \\
% \hline
% Text requests     & $4.5$M & $245$k & \begin{tabular}[c]{@{}l@{}}$182$k - $74.26$\%\\ $42$k - $17.42$\%\\ $20$k - $8.32$\%\end{tabular} & $3$ - $221$ - $27.57$ - $23$ - $54$                                                                 \\
% \hline
% Item descriptions & $4.5$M & $218$k & N/A                                                                                           & $9$ - $52$ - $20.68$ - $20$ - $25$                                                                 \\
% \hline
% \end{tabular}
% \caption{Data analysis of the WhisperD dataset.}
% \end{table*}

\begin{table}[h]
\small
\centering
\begin{tabular}{|l|l|l|}
\hline
     &  Text requests & \begin{tabular}[c]{@{}l@{}}Item\\ descriptions\end{tabular} \\
\hline
Number     & $4.5$M & $4.5$M  \\                    
\hline
Unique & $245$k & $218$k \\     
\hline
Sentiment (pos-neutr-neg) & $74.26$\% - $17.42$\% - $8.32$\% & N/A \\
\hline
Word length (min-max-avg) & $3$ - $221$ - $27.57$ & $9$ - $52$ - $20.68$ \\
\hline
Word length (50th-90th perc.) & $23$ - $54$ & $20$ - $25$  \\
\hline
\end{tabular}
\caption{Data analysis of the WhisperD dataset.}
\end{table}\label{tab:kaspian_data_an}

The main dataset used in this work, WhisperD, contains text requests from a personalized online fashion recommendation system, collected from real interactions between users and fashion stylists. We additionally consider adaptations of three open source datasets to demonstrate the broad applicability of our methodology. The statistics of the datasets are shown in Table \ref{tab:data_stats}.
\subsubsection{WhisperD}
The text requests in the dataset contain both positive preferences, e.g., \emph{Spring and summer wear}, as well as negative ones, e.g., \emph{No dresses, skirts}, and encompass a wide variety of writing styles.
Alongside every request in the dataset is also a collection of fashion stylists' item recommendations catering to the request. These item recommendations are categorized in three ways: items that the user decided to try on (TRYs), items that the user bought (KEEPs), and items that the user did not even try (NOTTRYs).
Table \ref{tab:kaspian_data_an} describes the details of the WhisperD dataset. All of the items associated with a user's request are chosen by a fashion stylist, and are therefore classified as positive examples. On the other hand, the negative examples are sampled randomly from the dataset.

\subsubsection{Open-source Datasets}
The open source datasets are obtained through slight modifications of the existing \textbf{Yelp} (restaurants domain) and \textbf{Amazon} (movies\&TV and clothing) datasets.
Whereas the original datasets contain customer reviews with binary sentiment annotations, we instead transform the datasets to match customer reviews to item product descriptions \cite{wang2022learning, kang2017visually}, aligning the datasets to the modality of WhisperD.
With this transformation, given a review and a set of possible targets, the task is to pair the review with the correct target description.
As described in \cite{wang2022learning, kang2017visually}, the content expressed through reviews can be interpreted as personal user preferences.

\subsubsection{Evaluation}
The evaluation measures used on the aforementioned datasets are the standard in recommender systems literature.
The main goal is to determine whether the items retrieved by the model satisfy the user's interest based on their initial query.
\begin{itemize}
    \item Precision@$k$ measures the number of recommended items in the top-$k$ that are relevant, 
    \item Recall@$k$ calculates the number of relevant products that are in the top-$k$ recommended ones,
    \item Normalized Discounted Cumulative Gain (NDCG) measures the relevancy of the item based on its position in the list of recommended products.
\end{itemize}
For precision@$k$ and recall@$k$ we chose $k \in \{1,2,3,4\}$ because of the varying number of relevant items for example.

\begin{table}[ht]
\small
\begin{tabular}{|l|lll|}
\hline
       Tasks & Train & Dev & Test \\
\hline
WhisperD   &   $1.58$M    &   $197$k  &   $197$k   \\
% Kaspian 40-60   &    $1.58$M   &  $197$k   &  $197$k    \\
Yelp            &  $11.3$M   &  $1.4$M   &   $1.4$M   \\
Amazon Clothing &   $2.22$M    &  $278$k   &  $278$k    \\
Amazon Movies   &   $13.52$M    &  $1.69$M   &   $1.69$M   \\
\hline
\end{tabular}
\caption{Statistics of all datasets used for this work.}
\label{tab:data_stats}
\end{table}
% 16.9M (clothing)
% 2.78M (movies)

\subsection{Results} \label{sec:results}
% \usepackage{multirow}
\begin{table*}[t]
\resizebox{16cm}{!}{
\begin{tabular}{|l|l|lllllllll|}
\hline
                                          &             & PREC@1 & PREC@2 & PREC@3 & PREC@4 & REC@1  & REC@2  & REC@3  & REC@4  & NDCG   \\
\hline
WhisperD                   & Random       & 0.5010 & 0.4667 & 0.4200 & 0.3697 & 0.2968 & 0.5250 & 0.6730 & 0.7571 & 0.6958 \\
                                          & OkapiBM25   &  0.5044      & 0.5069 & 0.5067 &   \textbf{0.5072}     &  0.0321      & 0.0646 & 0.0968  &    0.1293    & 0.7864 \\
                                          & Whisper    & 0.5376 & 0.4917 & 0.4343 & 0.377 & 0.3217 & 0.5515 & 0.6916 & 0.7678 & 0.712 \\
                                          & WLite       & \textbf{0.7633} & \textbf{0.6294} & \textbf{0.5111} & 0.4172 & \textbf{0.4707} & \textbf{0.6884} & \textbf{0.7817} & \textbf{0.8188} & \textbf{0.8088} \\
                                          
\hline
\hline
Yelp                                      & Whisper    & 0.8532 & 0.4316 & 0.289  & 0.2173 & \textbf{0.8473} & 0.8514 & 0.8524 & 0.8528 & 0.8533 \\
                                          & WLite       & \textbf{0.8554}   & \textbf{0.4377} & \textbf{0.2943} & \textbf{0.2218} & 0.8440 & \textbf{0.8527} & \textbf{0.8547} & \textbf{0.8555} & \textbf{0.8562} \\
\hline
\hline
Amazon movies\&TV                         & Whisper    & 0.7695 & 0.3881 & 0.2596 & 0.1951 & 0.7656 & 0.7683 & 0.7689 & 0.7692 & 0.7696 \\
                                          & WLite       & \textbf{0.7761} & \textbf{0.3953} & \textbf{0.2652} & \textbf{0.1996} & \textbf{0.7682} & \textbf{0.7746} & \textbf{0.7759} & \textbf{0.7765} & \textbf{0.7768} \\
\hline
\hline
Amazon clothes, shoes \& jewelry          & Whisper    & \textbf{0.8793} & 0.4516 & 0.3043 & 0.2298 & \textbf{0.8648} & \textbf{0.8743} & \textbf{0.8765} & \textbf{0.8776} & \textbf{0.8794} \\
                                          & WLite       & 0.8719 & \textbf{0.4587} & \textbf{0.3121}  & \textbf{0.2368} & 0.845  & 0.8637 & 0.8685 & 0.8705 & 0.8732 \\
\hline
\end{tabular}
}
\caption{Results table.}
\label{tab:results}
\end{table*}

Table \ref{tab:results} shows the results of the experiments on the previously presented datasets.
The \texttt{WhisperLite} model (\texttt{WLite}) substantially outperforms the other baselines on the WhisperD data on almost all metrics.
There is also a considerable gap between the \texttt{Whisper} and \texttt{WhisperLite} models, which only appears in the results on WhisperD.
One explanation is the WhisperD dataset consists of noisier text requests and label when compared to public datasets.
Higher noise settings could lead to a great degree of overfitting in \texttt{Whisper} model relative to the \texttt{WhisperLite} model, as the \texttt{Whisper} model optimizes over feature vectors for the text requests by users whereas the \texttt{WhisperLite} model does not.
One possible interpretation of such a difference could be that a model trained on generating embeddings for `generic' text requests, such as `\emph{Looking to update my wardrobe}' or `\emph{Going to Vegas next month}', diverts its ability to generate meaningful feature vectors for the description of the clothes that are paired with such requests.
Instead, \texttt{WLite} exploits the feature vectors generated by the pre-trained CLIP model to train only the two perceptrons with a classification objective function, which could explain the meaningful difference in performance on the WhisperD data.

On the open-source datasets the two models perform very similarly with a generally higher performance on all metrics compared to the results on WhisperD, hinting at the difficulty of the clothing recommendation task.
The results on the Yelp and Amazon clothing datasets are similar, whereas the performance on WhisperD is closer to the Amazon movies dataset.

\subsection{Human Evaluation} \label{sec:human_eval}
The evaluation compared the \texttt{WhisperLite} model with the \texttt{random} baseline through a user study on Amazon Mechanical Turk.
A total of $1500$ data points are randomly sampled from the outputs of the two respective models on the WhisperD dataset and submitted to the participants of the study for evaluation.
Each example is associated with $k$ recommended items, where $k \in \{3,5,7\}$.
In order to increase accountability and discard possible noise, every data point has been rated by $3$ unique workers.

The set up of this study is as follows: users are shown one text request and $k$ recommended products associated with it.
They then answer the question "\emph{How relevant are these items based on the text request?}" on a likert-scale from $1$ to $5$, where $1$ represents `Definitely irrelevant', $5$ means `Definitely relevant', and $3$ is left as the option for `Not sure'.
The studies for \texttt{WhisperLite} and \texttt{random} are designed to be separate and not as a comparison between the two, with the purpose of avoiding an introduction of bias on workers completing the tasks.
Furthermore, users were required specific qualifications to participate in this study and ensure an acceptable level of quality to the results, namely: being master workers, having completed and submitted at least $500$ HITs on MTurk with an acceptance rate greater than $80\%$, and being located in an English-speaking country.
Table \ref{tab:human_eval} shows the impact of both systems on real-world users.
We first classified `negative' ratings such as $1$ and $2$ as `$-1$', and `positive' ratings like $4$ and $5$ as `$+1$', whereas $0$ has been assigned to `neutral' judgements.

% One downside to this approach is that if all three workers provide completely different judgements to the same example, the agreement would still be $0.33$.

For all three values of $k$, where $k$ items are recommended given a specific request, users preferred the recommendations made by the \texttt{WhisperLite} model, with an average rating much higher than the recommendations from the \texttt{random} baseline.
The error bars computed on the average scores show the variability of the collected judgements from human judges.
Instead, the error bars for the \texttt{random} baseline indicate that the human ratings are less reliable than the ones collected for \texttt{WLite}, as they are bigger (or almost as big, for $k=7$) than the mean values.
This shows a net preference towards the items recommended by \texttt{WLite}.

\begin{table}[h]
\small
\centering
\begin{tabular}{|l|l|c|}
\hline
                     &         & Average \\
\hline
{$k=3$} & Random  & $0.31 \pm 0.37$   \\
                        & WLite   & $\textbf{0.52} \pm 0.28$    \\
\hline
\hline
{$k=5$} & Random  & $0.31 \pm 0.38 $\\
                        & WLite  & $\textbf{0.72} \pm 0.24 $ \\
\hline
\hline
{$k=7$} & Random  & $0.38 \pm 0.35 $ \\
                         & WLite   & $\textbf{0.59} \pm 0.27 $ \\
\hline
\end{tabular}
\caption{Analysis of the human evaluation of the \texttt{random} baseline and the \texttt{WLite} model.}
\label{tab:human_eval}
\end{table}

\section{Conclusion and Future Work}
In this work we describe recommending fashion items for a customer using a single text request. We propose two different approaches, Whisper and WhisperLite, that each address this session-based recommendation problem.
% Collaborative filtering makes use of past purchases from the users to propose sensible items.
While both approaches are competitive in several open source datasets spanning multiple domains, WhisperLite significantly outperforms all other baselines on the WhisperD dataset. We attribute the difference in model performance to the higher level of noise in the WhisperD dataset, where fine-tuning lower, language-model layers of model is highly sub-optimal, as it can confuse the system and negatively influence the generation of the features for the products (see Section \ref{sec:results}).
A meaningful extension of the proposed recommendation approach is to consider several other features like images of the clothing items or the body shape of the customer, in order to recommend items that could complement specific body shapes. Possible features that can also be included are user behavioral data from previous browsing sessions or historical data. We leave exploring the impact of these additional features as future work.

\bibliographystyle{ACM-Reference-Format}
\bibliography{custom}

% \appendix









\end{document}

%% file: arch.tex
\begin{tikzpicture}[
    % GLOBAL CFG
    font=\sf \large,
    % >=LaTeX,
    % Styles
    cell/.style={% For the main box
        rectangle, 
        rounded corners=5mm, 
        draw,
        very thick,
        },
    % operator/.style={%For operators like +  and  x
    %     circle,
    %     draw,
    %     inner sep=-0.5pt,
    %     minimum height =.2cm,
    %     },
    % function/.style={%For functions
    %     ellipse,
    %     draw,
    %     inner sep=1pt
    %     },
    ct/.style={% For external inputs and outputs
        rectangle,
        draw,
        line width = .75pt,
        minimum width=3cm,
        minimum height=3mm,
        inner sep=1pt,
        },
    gt/.style={% For internal inputs
        rectangle,
        draw,
        minimum width=1.5cm,
        minimum height=3mm,
        inner sep=1pt
        },
    mylabel/.style={% something new that I have learned
        font=\large\sffamily
        },
    mylabelout/.style={% something new that I have learned
        font=\large\sffamily
        },
    itemlabel/.style={% something new that I have learned
        font=\large\sffamily,
        color=red,
        },
    ArrowC1/.style={% Arrows with rounded corners
        thick,
        color=red
        },
    ArrowC2/.style={% Arrows with big rounded corners
        thick,
        color=red,
        style=dashed,
        },
    ArrowC3/.style={% Arrows with big rounded corners
        thick,
        },
    ]

%Start drawing the thing...    
    % Draw the cell: 
    \node [cell, minimum height=1.5cm, minimum width=14cm] (xlnet) at (-0.5,0){XLNet} ;
    \node [cell, minimum height=1cm, minimum width=6cm] (lstm1) at (-4,3){BiLSTM} ;
    \node [cell, minimum height=1cm, minimum width=6cm] (lstm2) at (3.5,3){BiLSTM} ;

    % Draw inputs named ibox#
    \node [gt] (ibox1) at (-6.25, 1.35) {};
    \node [gt] (ibox2) at (-4.75, 1.75) {};
    \node [gt] (ibox3) at (-3.5, 1.35) {};
    \node [gt] (ibox4) at (-2.5, 1.75) {};
    \node [gt] (ibox5) at (-1.5, 1.35) {};
    \node [ct] (output) at (-4, 4) {};
    
    \node [gt] (iboxitem1) at (1, 1.35) {};
    \node [gt] (iboxitem2) at (2.25, 1.75) {};
    \node [gt] (iboxitem3) at (3.5, 1.35) {};
    \node [gt] (iboxitem4) at (4.5, 1.75) {};
    \node [gt] (iboxitem5) at (6, 1.35) {};
    \node [ct] (itemoutput) at (3.5, 4) {};
    
    \node[mylabel](i) at (-6,-2.5) {I};
    \node[mylabel](need) at (-5.25,-2.5) {need};
    \node[mylabel](tops) at (-4.25,-2.52) {tops};
    \node[mylabel](and) at (-3,-2.5) {and};
    \node[mylabel](skirts) at (-1.75,-2.5) {skirts};
    \node[mylabel](enc1) at (-6,-1.65) {4};
    \node[mylabel](enc2) at (-5.25,-1.65) {67};
    \node[mylabel](enc3) at (-4.25,-1.652) {20495};
    \node[mylabel](enc4) at (-3,-1.65) {32};
    \node[mylabel](enc5) at (-1.75,-1.65) {45029};
    
    \node[mylabelout](sim) at (-0.5,4.75) {$cos(\theta)$};
    
    \node[itemlabel](woman) at (0.75,-2.52) {woman};
    \node[itemlabel](grey) at (1.75,-2.55) {grey};
    \node[itemlabel](crop) at (2.75,-2.55) {crop};
    \node[itemlabel](top) at (3.75,-2.52) {top};
    
    \node[itemlabel](encitem1) at (0.75,-1.65) {9};
    \node[itemlabel](encitem2) at (1.75,-1.65) {31};
    \node[itemlabel](encitem3) at (2.75,-1.652) {567};
    \node[itemlabel](encitem4) at (3.75,-1.65) {1045};

    % \node [gt] (ibox2) at (-1.5,-0.75) {$\sigma$};
    % \node [gt, minimum width=1cm] (ibox3) at (0,3) {Tanh};
    % \node [gt] (ibox4) at (0.5,-0.75) {$\sigma$};

   % Draw opérators   named mux# , add# and func#
    % \node [operator] (mux1) at (-2,1.5) {$\times$};
    % \node [operator] (add1) at (-0.5,1.5) {+};
    % \node [operator] (mux2) at (-0.5,0) {$\times$};
    % \node [operator] (mux3) at (1.5,0) {$\times$};
    % \node [function] (func1) at (1.5,0.75) {Tanh};

    % Draw External inputs? named as basis c,h,x
    % \node[ct, label={[mylabel]Cell}] (c) at (-4,1.5) {\empt{c}{t-1}};
    % \node[ct, label={[mylabel]Hidden}] (h) at (-4,-1.5) {\empt{h}{t-1}};
    % \node[ct, label={[mylabel]left:Input}] (x) at (-2.5,-3) {\empt{x}{t}};

    % Draw External outputs? named as basis c2,h2,x2
    % \node[ct, label={[mylabel]Label1}] (c2) at (4,1.5) {\empt{c}{t}};
    % \node[ct, label={[mylabel]Label2}] (h2) at (4,-1.5) {\empt{h}{t}};
    % \node[ct, label={[mylabel]left:Label3}] (x2) at (2.5,3) {\empt{h}{t}};

% Start connecting all.
    %Intersections and displacements are used. 
    % Drawing arrows    
    % \draw [ArrowC1] (c) -- (mux1) -- (add1) -- (c2);
    \draw [ArrowC3][-stealth] (xlnet) -- (ibox1);
    % \draw [ArrowC3][-stealth] (ibox1)-- (lstm1);
    % \draw [ArrowC3][-stealth] (xlnet) -- (ibox2);
    \draw [ArrowC3][-stealth] (ibox2)-- (lstm1);
    \draw [ArrowC3][-stealth] (xlnet) -- (ibox3);
    \draw [ArrowC3][-stealth] (ibox3)-- (lstm1);
    % \draw [ArrowC3][-stealth] (xlnet) -- (ibox4);
    \draw [ArrowC3][-stealth] (ibox4) -- (lstm1);
    \draw [ArrowC3][-stealth] (xlnet) -- (ibox5);
    % \draw [ArrowC3][-stealth] (ibox5) -- (lstm1);
    \draw [ArrowC3][-stealth] (lstm1) -- (output);
    \draw [ArrowC3][-stealth] (output) -- (sim);
    
    \draw [ArrowC1][-stealth] (xlnet) -- (iboxitem1);
    % \draw [ArrowC1][-stealth] (iboxitem1)-- (lstm2);
    % \draw [ArrowC1][-stealth] (xlnet) -- (iboxitem2);
    \draw [ArrowC1][-stealth] (iboxitem2)-- (lstm2);
    \draw [ArrowC1][-stealth] (xlnet) -- (iboxitem3);
    \draw [ArrowC1][-stealth] (iboxitem3)-- (lstm2);
    % \draw [ArrowC1][-stealth] (xlnet) -- (iboxitem4);
    \draw [ArrowC1][-stealth] (iboxitem4) -- (lstm2);
    \draw [ArrowC1][-stealth] (xlnet) -- (iboxitem5);
    % \draw [ArrowC1][-stealth] (iboxitem5) -- (lstm2);
    \draw [ArrowC1][-stealth] (lstm2) -- (itemoutput);
    \draw [ArrowC1][-stealth] (itemoutput) -- (sim);

    % Inputs
    \draw [ArrowC3][-stealth] (i) -- (enc1);
    \draw [ArrowC3][-stealth] (enc1) -- (xlnet);
    \draw [ArrowC3][-stealth] (need) -- (enc2);
    \draw [ArrowC3][-stealth] (enc2) -- (xlnet);
    \draw [ArrowC3][-stealth] (tops) -- (enc3);
    \draw [ArrowC3][-stealth] (enc3) -- (xlnet);
    \draw [ArrowC3][-stealth] (and) -- (enc4);
    \draw [ArrowC3][-stealth] (enc4) -- (xlnet);
    \draw [ArrowC3][-stealth] (skirts) -- (enc5);
    \draw [ArrowC3][-stealth] (enc5) -- (xlnet);
    
    \draw [ArrowC2][-stealth] (woman) -- (encitem1);
    \draw [ArrowC2][-stealth] (encitem1) -- (xlnet);
    \draw [ArrowC2][-stealth] (grey) -- (encitem2);
    \draw [ArrowC2][-stealth] (encitem2) -- (xlnet);
    \draw [ArrowC2][-stealth] (crop) -- (encitem3);
    \draw [ArrowC2][-stealth] (encitem3) -- (xlnet);
    \draw [ArrowC2][-stealth] (top) -- (encitem4);
    \draw [ArrowC2][-stealth] (encitem4) -- (xlnet);
    
    % \draw [ArrowC2] (h) -| (ibox4);
    % \draw [ArrowC1] (h -| ibox1)++(-0.5,0) -| (ibox1); 
    % \draw [ArrowC1] (h -| ibox2)++(-0.5,0) -| (ibox2);
    % \draw [ArrowC1] (h -| ibox3)++(-0.5,0) -| (ibox3);
    % \draw [ArrowC1] (x) -- (x |- h)-| (ibox3);

    % % Internal
    % \draw [->, ArrowC3] (xlnet) -- (ibox1);
    % \draw [->, ArrowC2] (ibox2) |- (mux2);
    % \draw [->, ArrowC2] (ibox3) -- (mux2);
    % \draw [->, ArrowC2] (ibox4) |- (mux3);
    % \draw [->, ArrowC2] (mux2) -- (add1);
    % \draw [->, ArrowC1] (add1 -| func1)++(-0.5,0) -| (func1);
    % \draw [->, ArrowC2] (func1) -- (mux3);

    % %Outputs
    % \draw [-, ArrowC2] (mux3) |- (h2);
    % \draw (c2 -| x2) ++(0,-0.1) coordinate (i1);
    % \draw [-, ArrowC2] (h2 -| x2)++(-0.5,0) -| (i1);
    % \draw [-, ArrowC2] (i1)++(0,0.2) -- (x2);

\end{tikzpicture}